\documentclass{article}
\usepackage{graphicx}
\usepackage{caption}
\usepackage{subcaption}
\usepackage{multicol}
\graphicspath{ {./images/} }

\usepackage{listings}
\usepackage{color}
\usepackage{float}

\usepackage{hyperref}
 \hypersetup{
     colorlinks=true,
     linkcolor=blue,
     filecolor=blue,
     citecolor = blue,      
     urlcolor=cyan,
     }

\definecolor{dkgreen}{rgb}{0,0.6,0}
\definecolor{gray}{rgb}{0.5,0.5,0.5}
\definecolor{mauve}{rgb}{0.58,0,0.82}

\usepackage[backend=biber,style=numeric,sorting=none]{biblatex}
\usepackage{geometry}
 \usepackage{braket}

 \usepackage{authblk}
\author[1]{Alex Linden}
\author[2]{Betül Gül}
\affil[1]{\textit{\small{Faculty of Mathematics and Computer Science, Jagiellonian University,  Kraków}}}
\affil[2]{\textit{\small{Faculty of Computer and Information Sciences, Sakarya University, Sakarya}}}

\title{Optimization of Postselection in Quantum Algorithms:\\ A Two-Way Quantum Computing Approach}

\date{}

\begin{document}

\maketitle{}

\small{\textbf{Abstract} - \textit{Postselection} is an operation that allows the selection of specific measurement outcomes. It serves as a powerful theoretical tool for enhancing the performance of existing quantum algorithms. Despite recent developments such as \textit{time reversal} in quantum measurements and IBM's \textit{mid-circuit measurements}, postselection continues to face significant challenges, most notably poor, often exponential, scaling. This study investigates how \textit{Two-Way Quantum Computing} (2WQC) offers potential solutions to these challenges. By introducing the concept of \textit{postparation} and enabling dynamic quantum state control, 2WQC has the potential to mitigate scaling issues and improve the practicality of postselection, thereby fostering advancements in the field of quantum algorithms.}\\

\begin{multicols}{2}[]

\section{Introduction}

\subsection{Postselection}

\textit{Postselection} is a procedure for discarding the runs of a quantum computation that yield a specific measurement outcome \cite{aaronson2005}. For example, it can be used as a projection operator $\ket{0}\bra{0}$, and while it does not affect $\ket{0}$, it discards all results $\ket{1}$ (similarly, if we implement postselection as an operator $\ket{1}\bra{1}$, it will discard all results $\ket{0}$ while retaining the $\ket{1}$ results.) It is important to note that postselection is not one established quantum operation but rather a procedure defined by its goal, i.e. conditioning specific measurement outcomes.

\subsection{Two-Way Quantum Computing}

\textit{One-Way Quantum Computing} (1WQC) exists as an analog to the standard quantum logic network. It utilizes one-qubit projective measurements on so called cluster states, highly-entangled multi-qubit states, which constitute the exclusive resource in this type of computation. \cite{Raussendorf_2002} 1WQC is unidirectional, since a cluster state's entanglement is irreversibly destroyed during measurement \cite{raussendorf2002computationalmodelunderlyingoneway}.\\

\textit{Two-Way Quantum Computing} (2WQC), introduced in \cite{duda2023}, functions much like 1WQC, while employing an approach based on CPT symmetry that allows for an introduction of a counterpart to state preparation $\ket{0}$, deemed \textit{postparation}. Physically, it could be implemented using EM impulses specific to quantum processors based on silicon quantum dots. The impulses could be reversed with regards to time ($V(t) \rightarrow V(-t)$) to perform postparation. Due to the change of direction, $\ket{0}$ would change to its symmetric analog $\bra{0}$, which would in turn induce a projection on one of its neighboring qubits. That projection would yield the same results as would an operator $\ket{0}\bra{0}$ applied to that qubit, and so could function as postselection. \cite{duda20243satsolvertwowayquantum}\\

This paper aims to examine 2WQC-based postselection, putting an emphasis on the possible improvements to existing methods of performing the procedure, its viability in specific algorithms and possible further avenues of research.

\section{Physical Viability of Postselection}

\subsection{Postselection and Weak measurements}

Physical implementations of postselection often rely on weak measurements \cite{kastner2017}. It’s a topic that has been studied extensively, with results confirming its viability in terms of compatibility to the formalism of quantum mechanics and experimental practicality. \cite{ashhab2009} 

The error-disturbance uncertainty (the limitation of measurement on two incompatible observables \cite{zhu2021}) is unaffected under postselection measurements which can facilitate reducing relative system error \cite{liu2019}.

\subsubsection{Enhancing Quantum Measurement Precision and Error Reduction with 2WQC}

State postparation is achieved through the time reversal of the electromagnetic (EM) impulses that prepare the quantum state \(|0\rangle\) \cite{noor2024}. This reversal enforces the state \(\langle 0|\) rather than selecting a measurement outcome via classical postselection methods. Unlike classical methods, which require repeated trials to obtain the desired state, postparation enforces the final state physically, which holds promise for improving the success rates of quantum algorithms under specific conditions \cite{noor2024}. In this way, the system’s final state is enforced during weak measurements, potentially leading to higher accuracy, although practical implementations still face significant challenges \cite{noor2024}.

Kastner's research emphasizes that, under appropriate conditions, weak measurements can provide results as precise as strong measurements, especially when the proper correlations are established during the measurement process \cite{kastner2017}. In this context, 2WQC has the potential to improve these correlations by introducing bidirectional control of the quantum states involved. This improvement arises from the enforcement of both the initial and final states during weak measurements, which theoretically enhances the accuracy of the results. This theoretical approach that may require further experimental validation to demonstrate its full potential across various quantum systems \cite{noor2024}.

For example, in the domain of computational problems like 3-SAT solvers, the concept of postparation might improve success rates by overcoming some of the limitations of traditional postselection methods \cite{duda20243satsolvertwowayquantum}. This suggests that postparation might be a useful tool for specific algorithmic applications where conventional postselection is inefficient due to low success rates, yet this is an area still requiring significant exploration.

In addition to improving precision, 2WQC also addresses the problem of error-disturbance uncertainty (EDR), a significant challenge in quantum measurement \cite{busch2014}. By enforcing both the initial and final states simultaneously, 2WQC reduces the uncertainty introduced by measurement disturbance. This dual enforcement mechanism potentially minimizes measurement errors and enhances the stability of weak measurement outcomes, though practical implementation may still involve overcoming scaling issues related to error rates \cite{noor2024}. The reduction in error bounds observed through 2WQC could, therefore, offer an incremental improvement in the precision of quantum measurements.

The mathematical expression of this precision improvement is captured through weak value calculations:

\[
\hat{A}_w = \frac{\langle \psi_f | \hat{A} | \psi_i \rangle}{\langle \psi_f | \psi_i \rangle}
\]
\cite{ahararonv1988}.

In this framework, 2WQC facilitates more precise adjustments to both the initial state (\(\psi_i\)) and the final state (\(\psi_f\)), allowing for more accurate measurement outcomes \cite{noor2024}. These adjustments are a core feature of the system, effectively forcing quantum states at both ends of the system to maintain accuracy and stability throughout the measurement process \cite{noor2024}.

Furthermore, 2WQC’s approach to EDR highlights its potential to lower the error bounds of the uncertainty principle while maintaining precision in weak measurement scenarios. The dual state enforcement offered by this method appears to reduce the impact of measurement disturbance, potentially leading to more reliable quantum systems. However, it is critical to acknowledge that these results, while theoretically sound, still require extensive validation in real-world quantum computing environments \cite{busch2014}.

Thus, while 2WQC holds significant promise for advancing quantum measurement techniques, particularly in the context of weak measurements, its full capabilities remain to be explored and proven in practical applications. Continued research is essential to confirm whether this method can consistently deliver improved success rates and enhanced measurement precision across a range of quantum computing tasks \cite{noor2024}.

\subsection{Time Reversal and Symmetry} \label{reversal}

Multiple different approaches have been taken to formalise time symmetry, or \textit{time reversal}, in quantum mechanics (e.g. \cite{aharonov1985}). Postselection has been proposed as a tool for discovery of new symmetries in quantum mechanical systems. \cite{rotello2023} Additionally, it has been observed to provide significant improvement in terms of accuracy of the data in related applications. \cite{rotello2023}

An operator rewinding quantum measurements has been described in \cite{hiromasa2022}:

\[ R((\ket{z}\bra{z} \otimes I^{\otimes n-1})\ket{\psi} \otimes \ket{\mathcal{D}})= \ket{\psi}. \]

A time-bidirectional state formalism, unifying the standard quantum mechanical formalism with postselection and the time-symmetrized two-state (density) vector formalism, which deals with postselected states, has been introduced in \cite{kiktenko2022}.

\section{Possible Approaches to Testing and Implementing Postselection}

\subsection{Hadamard Test}

The Hadamard test is an elementary quantum algorithm for finding an arbitrary expectation value using an ancilla qubit. 
The algorithm has many applications, e.g. error mitigation \cite{Polla_2023}, gradient estimation \cite{heidari2024efficientgradientestimationvariational} or computing inner products \cite{henderson2023addressingquantumsfineprint}.

\begin{figure}[H]
\centering
\includegraphics[scale=0.4]{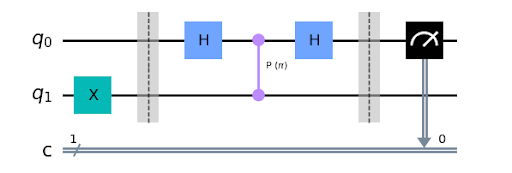}
\caption{A circuit for finding the real part $\textrm{Re}\bra{\psi}U\ket{\psi}$ in a Hadamard test.}
\end{figure}

\begin{figure}[H]
\centering
\includegraphics[scale=0.43]{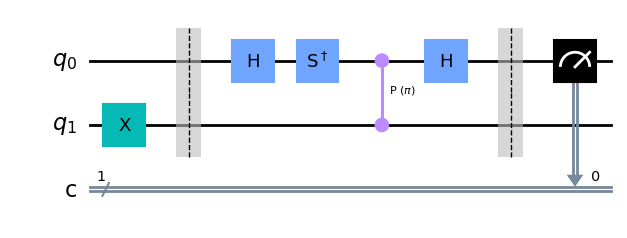}
\caption{A circuit for finding the imaginary part $\textrm{Im}\bra{\psi}U\ket{\psi}$ in a Hadamard test.}
\end{figure}

\subsubsection{Increasing Sensitivity with 2WQC}

The Hadamard test, shown in Figures 1 and 2, plays a critical role in various quantum computations by determining expected values. However, traditional implementations often require multiple repetitions to achieve the desired sensitivity due to uncertainties in quantum measurements \cite{henderson2023addressingquantumsfineprint}. The integration of 2WQC offers a novel theoretical solution to increase the sensitivity and efficiency of the Hadamard test by leveraging state postparation, as discussed in the 3-SAT solver approach \cite{duda20243satsolvertwowayquantum}.

2WQC theoretically introduces the concept of postparation, which allows for more controlled and precise quantum state evolution during both the preparation and measurement phases. This dynamic bidirectional control could potentially align the quantum state more precisely with the desired outcome, reducing variability and errors introduced by quantum noise \cite{duda20243satsolvertwowayquantum}. As a result, the 2WQC framework might enhance the sensitivity of the Hadamard test by providing more accurate calculations of expected values, thus potentially improving success rates in quantum computations \cite{duda20243satsolvertwowayquantum}.

\subsubsection{Potential for Reducing the Need for Repetition}

In traditional Hadamard tests, the need for multiple repetitions arises due to the probabilistic nature of quantum measurements \cite{henderson2023addressingquantumsfineprint}. However, 2WQC addresses this issue by allowing dynamic adjustment of the quantum state throughout the computation process. These adjustments could ensure that the state remains more consistent during both the initial and final stages of the computation, thereby potentially reducing errors commonly associated with quantum noise \cite{duda20243satsolvertwowayquantum}.

Additionally, 2WQC’s theoretical potential to solve \textbf{postBQP} problems (see \ref{complexity}) in a single run could allow the Hadamard test to yield more accurate results in a single trial \cite{noor2024}. By physically realizing the postselection process (e.g. through the reversal of EM pulses), results could be obtained much faster compared to classical methods \cite{noor2024}. This process may also enhance the overall efficiency of quantum computations by leveraging techniques such as state preparation and postparation.

\subsubsection{Theoretical Impacts on Computational Efficiency}

The integration of 2WQC with the Hadamard test offers significant theoretical improvements in computational efficiency \cite{Polla_2023}. By reducing error rates through bidirectional control, the need for additional quantum gates used for error correction could be minimized. This reduction may increase the speed of quantum algorithms such as gradient estimation or error mitigation while also improving their scalability \cite{Polla_2023}.

Furthermore, as theoretically demonstrated in the 3-SAT solver, the bidirectional enforcement of quantum states could enable the precise correction of errors mid-circuit \cite{duda20243satsolvertwowayquantum}. This feature of 2WQC could allow for faster and more accurate computations, making it a powerful tool for the future development of quantum algorithms. As discussed in \cite{duda20243satsolvertwowayquantum}, the inclusion of postparation and dynamic control over quantum states could significantly boost the reliability and efficiency of the Hadamard test.

\subsection{Quantum Phase Estimation}

The Quantum Phase estimation (QPE) is a subroutine for estimating the value of a unitary operator. Since it is based on so called \textit{black boxes} or \textit{oracles}, QPE is not a complete quantum algorithm but rather a module to be used as a part of another routine. \cite{nielsen00}

\begin{figure}[H]
\centering
\includegraphics[scale=0.22]{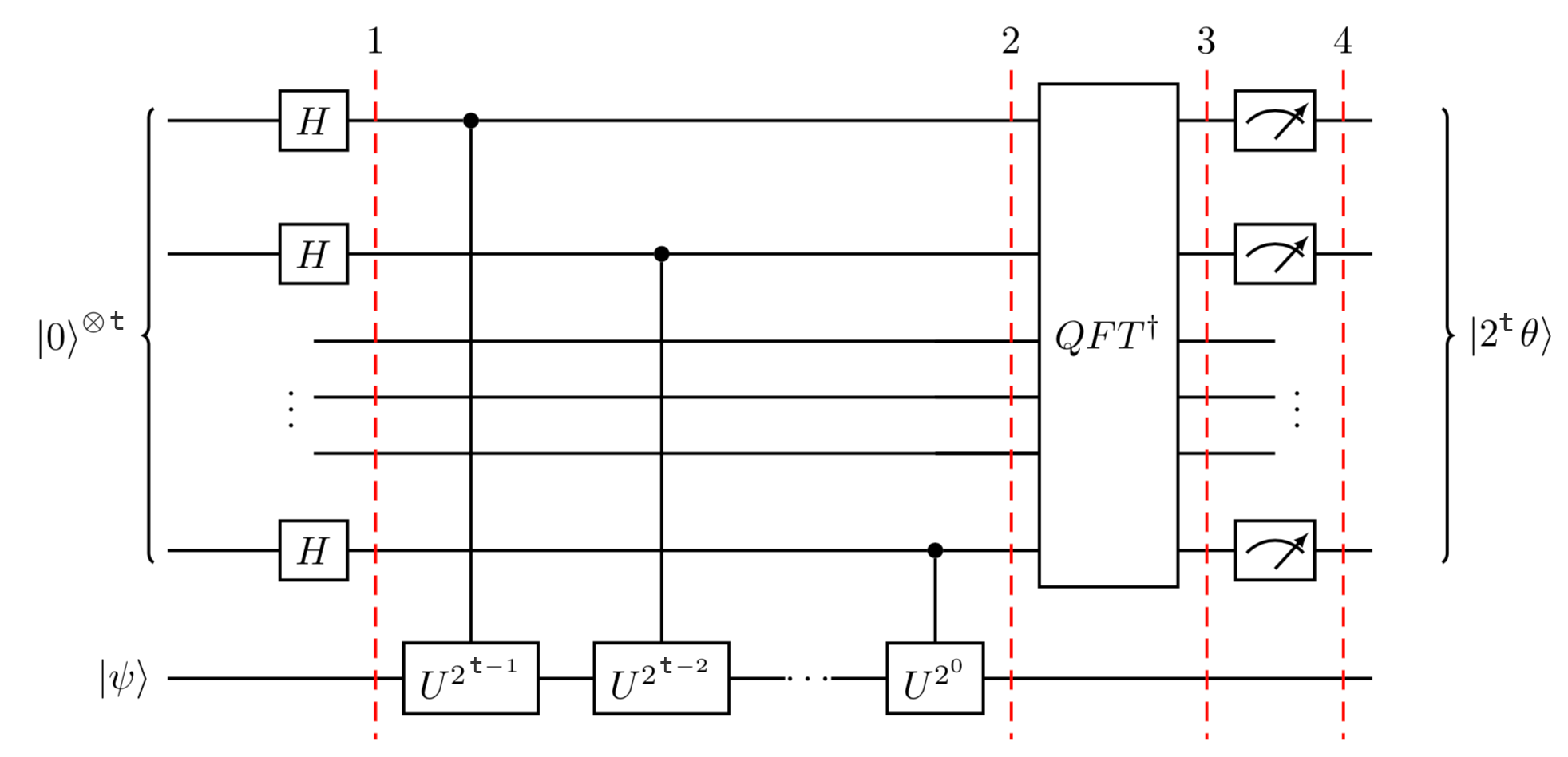}
\caption{The generalized circuit for quantum phase estimation. \cite{qiskitextbook2023}}
\end{figure}

Due to its versatility the subroutine is used in many important algorithms, Shor's algorithm for factoring large numbers \cite{Shor_1997} and the HHL algorithm for solving linear systems of equations \cite{Harrow_2009} among others. Additionally, it provides a basis for some quantum error correction algorithms. \cite{O_Brien_2021}\cite{PRXQuantum.4.040341}

\subsubsection{Optimizing Quantum Phase Estimation through 2WQC}

2WQC has the potential to address the exponential scaling issue associated with running postselection multiple times. Unlike traditional one-way quantum computers (1WQC), where state preparation occurs only at the beginning and measurement at the end, 2WQC actively controls both the initial and final states \cite{duda20243satsolvertwowayquantum}. In 2WQC, state postparation is the time-reversed analog of state preparation, achieved through the reversal of electromagnetic impulses. This symmetry allows for more precise state preparation and outcome enforcement, which may result in a more efficient implementation of Quantum Phase Estimation (QPE) \cite{duda20243satsolvertwowayquantum}. By reducing the need for repetitive postselection operations, 2WQC could improve the efficiency and accuracy of QPE-based algorithms, such as Shor’s algorithm for factoring and the HHL algorithm for solving linear systems of equations \cite{Shor_1997}. This improvement is due to the active control of final state enforcement, which mitigates the randomness associated with final measurements in standard quantum computing.

Postselection is often employed in quantum algorithms in conjunction with quantum error correction to mitigate the effects of time-dependent noise during computations \cite{O_Brien_2021}. However, emnploying this technique comes with scalability challenges due to the need for multiple runs to filter out undesirable outcomes. 2WQC, with its ability to dynamically control quantum states at both ends of the computation, may reduce the need for repeated postselection by stabilizing the computation and decreasing the reliance on repeated measurement corrections \cite{duda20243satsolvertwowayquantum}. Ensuring that quantum information remains consistent between the initial and final states, 2WQC can help to prevent phase drift during the computation, leading to more accurate phase estimates in fewer iterations \cite{Harrow_2009}.

Moreover, 2WQC’s ability to correct errors in real-time during computation without solely relying on postselection could further optimize the phase estimation process. By enforcing correct quantum states throughout the circuit, 2WQC stabilizes the computation, reducing the need for repeated trials in QPE \cite{noor2024}. The mid-circuit error correction mechanism inherent in 2WQC, discussed in the context of the 3-SAT solver, illustrates how dynamically enforced correct states can lead to higher accuracy and reduced computational overhead \cite{duda20243satsolvertwowayquantum}.

Another advantage of integrating 2WQC into QPE is the potential reduction in the number of qubits required for computation. Typically, achieving high accuracy in QPE demands a significant number of qubits and quantum gates. However, by reducing reliance on postselection, 2WQC can optimize quantum circuit design, decreasing the number of required qubits and making the overall computation more resource-efficient \cite{Mart_n_L_pez_2012}. This resource optimization is particularly important for scaling quantum algorithms to larger problems, where qubit and gate limitations pose significant challenges.

Finally, the no-cloning theorem remains valid within the 2WQC framework, ensuring that quantum information is preserved throughout the computation without violating fundamental quantum principles \cite{noor2024}. This preservation is critical for phase estimation algorithms, where precise control over quantum states is necessary to prevent information degradation. By offering enhanced control over quantum states, 2WQC can lead to more reliable phase estimation, even in the presence of quantum noise, making it a powerful tool for improving the performance of quantum algorithms that rely on QPE \cite{O_Brien_2021}.

\subsection{Shor's Algorithm}

Shor’s algorithm is a flagship quantum algorithm for factoring numbers. It allows for superpolynomial speedup over its classical counterpart. Experimental implementations have been performed using different methods, although only on relatively small numbers (e.g. 21.) \cite{Mart_n_L_pez_2012}\cite{Skosana_2021}

One of the main hurdles to overcome while implementing Shor’s algorithm is the large resource requirement
that scales poorly with the number that’s meant to be factored. To address this, a recycling method has been introduced that utilizes postselection in the form of a CNOT gate. Due to reducing the size of the register from $n$ qubits to one that is recycled $n$ times, the number of all qubits needed was three times less than in the original version of the algorithm. \cite{Mart_n_L_pez_2012}

The operation of postselection needs to be performed during the circuit execution and so it constitutes a
challenge in its experimental implementation. IBM proposed a solution to this by introducing dynamic circuits, allowing for mid-circuit measurements. \cite{nation_2021_how}

Using postselection comes with its own drawbacks. It cannot be used with some approaches to implementing Shor's algorithm. For example, in \cite{Skosana_2021} the number of measurements would scale as $n!$ since a measurement basis has to be known for each qubit.

\subsubsection{Two-Way Quantum Computing (2WQC) and Postselection in Shor’s Algorithm}

In some experimental implementations of Shor’s algorithm, postselection and qubit recycling are used to reduce resource requirements, although these techniques are not part of the standard algorithm \cite{Mart_n_L_pez_2012}. The specific use of postselection in the algorithm typically involves conditional operations during quantum circuits, such as in the mid-circuit measurement and reset of qubits, allowing for more efficient resource usage during computation. Integrating Two-Way Quantum Computing (2WQC) offers an alternative approach that could further increase efficiency. In particular, state postparation, the time-reversed counterpart to state preparation, provides a novel way to enhance postselection by increasing the likelihood of success without needing to know the measurement results beforehand \cite{duda2023}.

\subsubsection{Postparation and Improving the Success Rate}

In traditional \textit{postselection}, the success rate of the system often scales poorly because many measurement outcomes that do not meet the desired criteria are discarded. This leads to an exponential decline in efficiency for larger quantum systems, especially when Shor’s algorithm is required to factor larger numbers \cite{Mart_n_L_pez_2012}. However, 2WQC addresses this limitation by introduces postparation, which enforces the final state of the quantum system similarly to how state preparation enforces the initial state \cite{duda2023}. This mechanism reduces the number of discarded measurements and enhances the overall efficiency of quantum computations.

In the context of Shor’s algorithm, replacing the traditional postselection process with postparation in 2WQC could significantly improve the success rate. Instead of performing multiple postselections, postparation could be applied to physically enforce the result, potentially improving the success rate and mitigating the exponential scaling issues associated with postselection \cite{duda2023}.

\subsubsection{Qubit Recycling and 2WQC Integration}

Qubit recycling, as discussed in the experimental realization of Shor’s algorithm, is one method used to reduce the number of qubits required by resetting and reusing qubits during computation \cite{Mart_n_L_pez_2012, Skosana_2021}. This technique has been successfully demonstrated in small quantum systems, such as experimentally factoring the number 21 \cite{Skosana_2021}. However, this method is constrained by the need to perform mid-circuit measurements and resets, as each qubit must be measured and recycled during the computation process \cite{nation_2021_how}.

The integration of 2WQC into Shor’s algorithm offers a potential improvement in this scenario. By utilizing state postparation, the system no longer needs to know the measurement outcomes before resetting the qubits. Instead, postparation enforces the final state directly, allowing for qubit recycling without requiring prior knowledge of the measurement basis. This integration could streamline the implementation of Shor’s algorithm, reducing the number of required qubits and potentially increasing the overall success rate of the algorithm \cite{duda2023, duda20243satsolvertwowayquantum, Skosana_2021}.

\subsubsection{Potential Advantages and Applications}

The integration of 2WQC with qubit recycling could offer several advantages for Shor’s algorithm. By combining qubit recycling with postparation, fewer qubits would be needed, and the overall computation could become more efficient \cite{duda2023, duda20243satsolvertwowayquantum, Skosana_2021}. The success rate could be improved by enforcing the final state through postparation, which mitigates the poor scaling associated with postselection \cite{duda2023, Skosana_2021}. Additionally, enforcing the result through postparation could reduce the impact of errors, particularly in noisy quantum systems \cite{duda2023, Skosana_2021}. This approach could be extended beyond Shor’s algorithm to other quantum algorithms that rely on postselection and qubit recycling. The increased success rates and reduced qubit requirements could make more complex quantum computations feasible on near-term quantum devices \cite{duda2023, Skosana_2021}.

\subsection{Eigenvalue Finding}

Because the exponential scaling resulting from using postselection is described as a bottleneck in the eigenvalue finding algorithm, there is potential for improvement when utilising 2WQC. \cite{nghiem2023}\cite{duda2023}

\subsection{State Reconstruction}

State reconstruction is an example of utilising postselection that has not yet been extensively studied (mathematical description in \cite{chowdhury2023}.)

\subsection{QML Algorithms Utilising Postselection}

There exist multiple algorithms in the realm of quantum machine learning using postselection, such as variational quantum anstatz \cite{zhang2024, aharonov1985} or the quantum circuit Born machine (QCBM) \cite{zhu2011}.

\section{Possible Applications for Quantum Error Correction}

Postselection has found many applications in quantum error mitigation and correction.

The Accreditation protocol with postselection, providing error mitigation for time-dependent noise behaviours, has been tested experimentally \cite{mezher2021} with promising results. Since it requires multiple runs of the computation to perform postselection, it might be possible to improve upon it using only one run of 2WQC's postparation.

The probabilistic quantum error correction procedure, has been described and examined in \cite{kukulski2022}. Postselection, used in the protocol, acts as a classical post-processing to determine whether the encoded information has been successfully restored.

Postselected error correction \cite{Knill_2005} has been shown to significantly outperform regular quantum error correction. In the case of a surface code implementation with distance-four stabilizer measurement sequences accumulates, the improvement manifests in 25 times less accumulated errors than its counterpart \cite{prabhu2021}. Knill's postselection scheme, based on concatenated error-correcting codes, has been proven to have the highest known error threshold for two-dimensional concatenated codes. Similarly, it has been demonstrated that  topological stabilizer codes with qubit flagging can surpass a pseudo-threshold for repeated logical measurements \cite{PhysRevLett.128.110504}.

\cite{lai2023} Using postselection on syndrome outcomes for even distance code effectively increases the code's distance from $d$ to $d+1$ since errors of weight $\frac{d}{2}$ do not cause logical errors using that technique \cite{dasilva2024demonstrationlogicalqubitsrepeated, prabhu2021}.

The implementation of like procedures on a 2-way quantum computer could increase their viability and practicality. In some cases it could even pose a theoretical improvement through a reduction in resources needed to perform the computation.

\section{The Complexity Theory Approach} \label{complexity}

The notion of a PostBQP complexity class (Postselected Bounded-Error Quantum Polynomial-Time) was first introduced in \cite{aaronson2005}, where it is defined as the group of all problems that can be solved in polynomial time by a quantum computer with postselection capabilities. 

It has been shown that $\mathrm{\textbf{NP}} \subseteq \mathrm{\textbf{PostBQP}} \subseteq \mathrm{\textbf{PP}}$. \cite{aaronson2005limitsefficientcomputationphysical} Notably, $\mathrm{\textbf{PostBQP}} = \mathrm{\textbf{PP}}$ implies that postselection could help solve problems from $\mathrm{\textbf{NP}}$, since $\mathrm{\textbf{NP}} \subset \mathrm{\textbf{PP}}$. \cite{doi:10.1137/0206049}

\begin{figure}[H]
\centering
\includegraphics[scale=0.3]{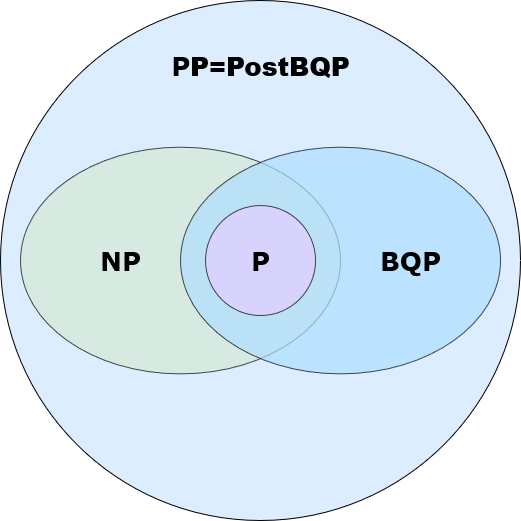}
\caption{Relation between the complexity classes \textbf{P}, \textbf{NP}, \textbf{BQP}, \textbf{PP} and \textbf{PostBQP}}
\end{figure}

As discussed in \ref{reversal}, time reversal and symmetry have been studied extensively. In particular, a new complexity class \textbf{RwBQP} has been introduced that includes all problems solvable with quantum circuits of polynomial size and  a polynomial number of operators rewinding quantum measurements. Additionally, it has been proven that $\mathrm{\textbf{BQP}} \subseteq \mathrm{\textbf{RwBQP}} \subseteq \mathrm{\textbf{PSPACE}}$ \cite{hiromasa2022}.

There exists a direct link between $\mathrm{\textbf{RwBQP}}$ and $\mathrm{\textbf{PostBQP}}$, since the equality $\mathrm{\textbf{PostBQP}} = \mathrm{\textbf{PP}}$ holds even if instead of standard quantum computers equipped with postselection, the rewindable quantum computation model is used, where the rewinding operators are replaced with postselection \cite{hiromasa2022}.

Since 2WQC can similarly employ bidirectionality to simulate postselection, an obvious question arises about a complexity class containing the set of problems solvable with a two-way quantum computer, say, \textbf{2WBQP} (2-Way Bounded-Error Quantum Polynomial-Time.) An analysis of its relationships to other complexity classes could help in understanding the more precise nature of 2WQC, along with problems it would be capable of solving.

If \textbf{2WBQP} is equal to \textbf{PostBQP}, then 2WQC would be capable of solving problems from \textbf{NP}.  Moreover, it has been suggested that using postselection and 2WQC could help solve \textbf{NP-complete} problems. For example, there exists a protocol for solving the k-colouring problem \cite{karacsony2023} using postselection, and a 2WQC-based approach for solving 3-SAT (the boolean satisfiability problem) \cite{duda20243satsolvertwowayquantum}.

\section{Conclusions}

Postselection is a procedure that has been used in numerous algorithms as a means to conditionally reduce the number of runs of a quantum computation. Current research examines this operation both in the theoretical and physical sense. Although it has been widely used in quantum computing and quantum information science, its physical viability is still being debated.

The novel 2-way quantum computing architecture employs time bidirectionality which enables the introduction of postparation - the time-symmetric counterpart to state preparation. It has potential to not only bridge the gap of physical implementation of postselection, but also improve upon it, for example reducing the number of needed runs of a computation.

In the paper, multiple possible approaches were analyzed. The physical time symmetry approach to implementing postselection relates to the CPT symmetry - a concept on which 2WQC heavily relies upon. Some flagship quantum algorithms, such as the Hadamard test, Quantum Phase Estimation or Shor's algorithm could potentially be improved by introducing postparation-based postselection. Additionally, there exist multiple quantum error-correction (QEC) schemes utilizing postselection - the introduction of a physically viable way of executing them would imply significant advances in the field of QEC. Finally, the analysis of a newly introduced \textbf{2WBQP} class in terms of its relations to existing classes (especially \textbf{PostBQP} and \textbf{RwBQP}) could help define the precise nature of problems solvable using a 2-way quantum computer.

\end{multicols}
\printbibliography

@book{nielsen00,
  author = {Nielsen, Michael A. and Chuang, Isaac L.},
  publisher = {Cambridge University Press},
  title = {Quantum Computation and Quantum Information},
  year = 2000
}

@article{kastner2017,
	author = {Ruth Kastner},
	doi = {10.1007/s10701-017-0085-4},
	journal = {Foundations of Physics},
	number = {5},
	pages = {697--707},
	publisher = {Springer Us},
	title = {Demystifying Weak Measurements},
	volume = {47},
	year = {2017}
}

@article{ashhab2009,
  author = {S. Ashhab and Franco Nori},
  title = {How the result of a measurement of a component of the spin of a spin-1/2 particle can turn out to be 100 without using weak measurements},
  journal = {arXiv},
  year = {2009},
  url = {https://arxiv.org/abs/0907.4823}
}

@article{zhu2011,
  title = {Quantum measurements with preselection and postselection},
  author = {Zhu, Xuanmin and Zhang, Yuxiang and Pang, Shengshi and Qiao, Chang and Liu, Quanhui and Wu, Shengjun},
  journal = {Phys. Rev. A},
  volume = {84},
  issue = {5},
  pages = {052111},
  numpages = {8},
  year = {2011},
  month = {Nov},
  publisher = {American Physical Society},
  doi = {10.1103/PhysRevA.84.052111},
  url = {https://link.aps.org/doi/10.1103/PhysRevA.84.052111}
}

@article{noor2024,
  author = {M. Noor and J. Duda},
  title = {No-cloning theorem for 2WQC and postselection},
  journal = {arXiv},
  year = {2024},
  url = {https://arxiv.org/abs/2407.15623}
}

@article{ahararonv1988,
  title={How the result of a measurement of a component of the spin of a spin-1/2 particle can turn out to be 100},
  author={Aharonov, Yakir and Albert, David Z. and Vaidman, Lev},
  journal={Phys. Rev. Lett.},
  volume={60},
  pages={1351-1354},
  year={1988}
}

@article{duda2023,
  author = {Jarek Duda},
  title = {Two-way quantum computers adding CPT analog of state preparation},
  journal = {arXiv},
  year = {2023},
  url = {https://arxiv.org/abs/2308.13522}
}

@article{busch2014,
  author = {Busch, Paul and Lahti, Pekka and Werner, Reinhard F.},
  title = {Colloquium: Quantum root-mean-square error and measurement uncertainty relations},
  journal = {Rev. Mod. Phys.},
  volume = {86},
  issue = {4},
  pages = {1261--1281},
  numpages = {0},
  year = {2014},
  month = {Dec},
  publisher = {American Physical Society},
  doi = {10.1103/RevModPhys.86.1261},
  url = {https://link.aps.org/doi/10.1103/RevModPhys.86.1261}
}

@article{liu2019,
author = {Yang Liu and Haijun Kang and Dongmei Han and Xiaolong Su and Kunchi Peng},
journal = {Photon. Res.},
number = {11},
pages = {A56--A60},
publisher = {Optica Publishing Group},
title = {Experimental test of error-disturbance uncertainty relation with continuous variables},
volume = {7},
month = {Nov},
year = {2019},
url = {https://opg.optica.org/prj/abstract.cfm?URI=prj-7-11-A56},
doi = {10.1364/PRJ.7.000A56},
}

@article{rotello2023,
  title = {Automated detection of symmetry-protected subspaces in quantum simulations},
  author = {Rotello, Caleb and Jones, Eric B. and Graf, Peter and Kapit, Eliot},
  journal = {Phys. Rev. Res.},
  volume = {5},
  issue = {3},
  pages = {033082},
  numpages = {22},
  year = {2023},
  month = {Aug},
  publisher = {American Physical Society},
  doi = {10.1103/PhysRevResearch.5.033082},
  url = {https://link.aps.org/doi/10.1103/PhysRevResearch.5.033082}
}

@InProceedings{hiromasa2022,
  author =	{Hiromasa, Ryo and Mizutani, Akihiro and Takeuchi, Yuki and Tani, Seiichiro},
  title =	{{Rewindable Quantum Computation and Its Equivalence to Cloning and Adaptive Postselection}},
  booktitle =	{18th Conference on the Theory of Quantum Computation, Communication and Cryptography (TQC 2023)},
  pages =	{9:1--9:23},
  series =	{Leibniz International Proceedings in Informatics (LIPIcs)},
  ISBN =	{978-3-95977-283-9},
  ISSN =	{1868-8969},
  year =	{2023},
  volume =	{266},
  editor =	{Fawzi, Omar and Walter, Michael},
  publisher =	{Schloss Dagstuhl -- Leibniz-Zentrum f{\"u}r Informatik},
  address =	{Dagstuhl, Germany},
  URL =		{https://drops.dagstuhl.de/entities/document/10.4230/LIPIcs.TQC.2023.9},
  URN =		{urn:nbn:de:0030-drops-183193},
  doi =		{10.4230/LIPIcs.TQC.2023.9}
}

@article{kiktenko2022,
  title = {Exploring postselection-induced quantum phenomena with time-bidirectional state formalism},
  author = {Kiktenko, Evgeniy O.},
  journal = {Phys. Rev. A},
  volume = {107},
  issue = {3},
  pages = {032419},
  numpages = {14},
  year = {2023},
  month = {Mar},
  publisher = {American Physical Society},
  doi = {10.1103/PhysRevA.107.032419},
  url = {https://link.aps.org/doi/10.1103/PhysRevA.107.032419}
}

@article{nghiem2023,
  author = {Nhat A. Nghiem and Tzu-Chieh We},
  title = {Improved Quantum Algorithms for Eigenvalues Finding and Gradient Descent},
  journal = {arXiv},
  year = {2023},
  url = {https://arxiv.org/pdf/2312.14786}
}

@misc{chowdhury2023,
  author = {Anirban Ch Narayan Chowdhury},
  title = {Quantum measurements with postselection},
  year = {2023},
  url = {http://dr.iiserpune.ac.in:8080/xmlui/handle/123456789/255}
}

@article{mezher2021,
  title = {Mitigating errors by quantum verification and postselection},
  author = {Mezher, Rawad and Mills, James and Kashefi, Elham},
  journal = {Phys. Rev. A},
  volume = {105},
  issue = {5},
  pages = {052608},
  numpages = {13},
  year = {2022},
  month = {May},
  publisher = {American Physical Society},
  doi = {10.1103/PhysRevA.105.052608},
  url = {https://link.aps.org/doi/10.1103/PhysRevA.105.052608}
}

@ARTICLE{kukulski2022,
  author={Kukulski, Ryszard and Pawela, Łukasz and Puchała, Zbigniew},
  journal={IEEE Transactions on Information Theory}, 
  title={On the Probabilistic Quantum Error Correction}, 
  year={2023},
  volume={69},
  number={7},
  pages={4620-4640},
  doi={10.1109/TIT.2023.3254054}}

@article{prabhu2021,
  title = {Distance-four quantum codes with combined postselection and error correction},
  author = {Prabhu, Prithviraj and Reichardt, Ben W.},
  journal = {Phys. Rev. A},
  volume = {110},
  issue = {1},
  pages = {012419},
  numpages = {13},
  year = {2024},
  month = {Jul},
  publisher = {American Physical Society},
  doi = {10.1103/PhysRevA.110.012419},
  url = {https://link.aps.org/doi/10.1103/PhysRevA.110.012419}
}

@article{lai2023,
  author = {Ching-Yi Lai and Gerardo Paz and Martin Suchara and Todd A. Brun},
  title = {Performance and error analysis of Knill’s postselection scheme in a two-dimensional architecture},
  journal = {Semantic Scholar},
  year = {2023},
  url = {https://www.semanticscholar.org/reader/8637513c95315cda7addfbbf4ac92ff482c3d1bb}
}

@article{aaronson2005,
  author = {Scott Aaronson},
  title = {Quantum Computing, Postselection, and Probabilistic Polynomial-Time},
  journal = {arXiv},
  year = {2005},
  url = {https://www.scottaaronson.com/papers/pp.pdf}
}

@article{karacsony2023,
  author = {M. Karácsony et al.},
  title = {Efficient qudit based scheme for photonic quantum computing},
  journal = {SciPost},
  year = {2023},
  url = {https://scipost.org/preprints/scipost_202304_00012v2/}
}

@article{zhang2024,
  author = {Shi-Xin Zhang and Jiaqi Miao and Chang-Yu Hsieh},
  title = {Variational post-selection for ground states and thermal states simulation},
  journal = {arXiv},
  year = {2024},
  url = {https://arxiv.org/pdf/2402.07605}
}

@article{zhu2021,
  title = {Generative quantum learning of joint probability distribution functions},
  author = {Zhu, Elton Yechao and Johri, Sonika and Bacon, Dave and Esencan, Mert and Kim, Jungsang and Muir, Mark and Murgai, Nikhil and Nguyen, Jason and Pisenti, Neal and Schouela, Adam and Sosnova, Ksenia and Wright, Ken},
  journal = {Phys. Rev. Res.},
  volume = {4},
  issue = {4},
  pages = {043092},
  numpages = {17},
  year = {2022},
  month = {Nov},
  publisher = {American Physical Society},
  doi = {10.1103/PhysRevResearch.4.043092},
  url = {https://link.aps.org/doi/10.1103/PhysRevResearch.4.043092}
}

@article{aharonov1985,
  author = {Yakir Aharonov et al.},
  title = {Time Symmetry in the Process of Quantum Measurement},
  journal = {CMSR Rutgers},
  year = {1985},
  url = {https://cmsr.rutgers.edu/images/people/lebowitz_joel/publications/jll.pub_34.pdf}
}

@article{Mart_n_L_pez_2012,
   title={Experimental realization of Shor’s quantum factoring algorithm using qubit recycling},
   volume={6},
   ISSN={1749-4893},
   url={http://dx.doi.org/10.1038/nphoton.2012.259},
   DOI={10.1038/nphoton.2012.259},
   number={11},
   journal={Nature Photonics},
   publisher={Springer Science and Business Media LLC},
   author={Martín-López, Enrique and Laing, Anthony and Lawson, Thomas and Alvarez, Roberto and Zhou, Xiao-Qi and O’Brien, Jeremy L.},
   year={2012},
   month=oct, pages={773–776} }

@article{Skosana_2021,
   title={Demonstration of Shor’s factoring algorithm for N = 21 on IBM quantum processors},
   volume={11},
   ISSN={2045-2322},
   url={http://dx.doi.org/10.1038/s41598-021-95973-w},
   DOI={10.1038/s41598-021-95973-w},
   number={1},
   journal={Scientific Reports},
   publisher={Springer Science and Business Media LLC},
   author={Skosana, Unathi and Tame, Mark},
   year={2021},
   month=aug }

@misc{nation_2021_how,
  author = {Nation, Paul},
  month = {02},
  title = {How to measure and reset a qubit in the middle of a circuit execution | IBM Quantum Computing Blog},
  url = {https://www.ibm.com/quantum/blog/quantum-mid-circuit-measurement},
  urldate = {2024-07-23},
  year = {2021},
  organization = {ibm.com}
}

@misc{heidari2024efficientgradientestimationvariational,
      title={Efficient Gradient Estimation of Variational Quantum Circuits with Lie Algebraic Symmetries}, 
      author={Mohsen Heidari and Masih Mozakka and Wojciech Szpankowski},
      year={2024},
      eprint={2404.05108},
      archivePrefix={arXiv},
      primaryClass={quant-ph},
      url={https://arxiv.org/abs/2404.05108}, 
}

@article{Polla_2023,
   title={Optimizing the information extracted by a single qubit measurement},
   volume={108},
   ISSN={2469-9934},
   url={http://dx.doi.org/10.1103/PhysRevA.108.012403},
   DOI={10.1103/physreva.108.012403},
   number={1},
   journal={Physical Review A},
   publisher={American Physical Society (APS)},
   author={Polla, Stefano and Anselmetti, Gian-Luca R. and O’Brien, Thomas E.},
   year={2023},
   month=jul }

@misc{henderson2023addressingquantumsfineprint,
      title={Addressing Quantum's "Fine Print": State Preparation and Information Extraction for Quantum Algorithms and Geologic Fracture Networks}, 
      author={Jessie M. Henderson and John Kath and John K. Golden and Allon G. Percus and Daniel O'Malley},
      year={2023},
      eprint={2310.02479},
      archivePrefix={arXiv},
      primaryClass={quant-ph},
      url={https://arxiv.org/abs/2310.02479}, 
}

@article{Harrow_2009,
   title={Quantum Algorithm for Linear Systems of Equations},
   volume={103},
   ISSN={1079-7114},
   url={http://dx.doi.org/10.1103/PhysRevLett.103.150502},
   DOI={10.1103/physrevlett.103.150502},
   number={15},
   journal={Physical Review Letters},
   publisher={American Physical Society (APS)},
   author={Harrow, Aram W. and Hassidim, Avinatan and Lloyd, Seth},
   year={2009},
   month=oct }

@article{Shor_1997,
   title={Polynomial-Time Algorithms for Prime Factorization and Discrete Logarithms on a Quantum Computer},
   volume={26},
   ISSN={1095-7111},
   url={http://dx.doi.org/10.1137/S0097539795293172},
   DOI={10.1137/s0097539795293172},
   number={5},
   journal={SIAM Journal on Computing},
   publisher={Society for Industrial & Applied Mathematics (SIAM)},
   author={Shor, Peter W.},
   year={1997},
   month=oct, pages={1484–1509} }

@article{O_Brien_2021,
   title={Error Mitigation via Verified Phase Estimation},
   volume={2},
   ISSN={2691-3399},
   url={http://dx.doi.org/10.1103/PRXQuantum.2.020317},
   DOI={10.1103/prxquantum.2.020317},
   number={2},
   journal={PRX Quantum},
   publisher={American Physical Society (APS)},
   author={O’Brien, Thomas E. and Polla, Stefano and Rubin, Nicholas C. and Huggins, William J. and McArdle, Sam and Boixo, Sergio and McClean, Jarrod R. and Babbush, Ryan},
   year={2021},
   month=may }

@article{PRXQuantum.4.040341,
  title = {Statistical Phase Estimation and Error Mitigation on a Superconducting Quantum Processor},
  author = {Blunt, Nick S. and Caune, Laura and Izs\'ak, R\'obert and Campbell, Earl T. and Holzmann, Nicole},
  journal = {PRX Quantum},
  volume = {4},
  issue = {4},
  pages = {040341},
  numpages = {23},
  year = {2023},
  month = {Dec},
  publisher = {American Physical Society},
  doi = {10.1103/PRXQuantum.4.040341},
  url = {https://link.aps.org/doi/10.1103/PRXQuantum.4.040341}
}

@book{qiskitextbook2023,   
    author = {various authors},   
    year = {2023},   
    title = {Qiskit Textbook},   
    publisher = {Github},   
    url = {https://github.com/Qiskit/textbook}, 
}

@article{Raussendorf_2002,
   title={The one-way quantum computer--a non-network model of quantum computation},
   volume={49},
   ISSN={1362-3044},
   url={http://dx.doi.org/10.1080/09500340110107487},
   DOI={10.1080/09500340110107487},
   number={8},
   journal={Journal of Modern Optics},
   publisher={Informa UK Limited},
   author={Raussendorf, Robert and Browne, Daniel and Briegel, Hans},
   year={2002},
   month=jul, pages={1299–1306} }

@misc{raussendorf2002computationalmodelunderlyingoneway,
      title={Computational model underlying the one-way quantum computer}, 
      author={Robert Raussendorf and Hans Briegel},
      year={2002},
      eprint={quant-ph/0108067},
      archivePrefix={arXiv},
      primaryClass={quant-ph},
      url={https://arxiv.org/abs/quant-ph/0108067}, 
}

@misc{duda20243satsolvertwowayquantum,
      title={3-SAT solver for two-way quantum computers}, 
      author={Jarek Duda},
      year={2024},
      eprint={2408.05812},
      archivePrefix={arXiv},
      primaryClass={physics.gen-ph},
      url={https://arxiv.org/abs/2408.05812}, 
}

@misc{aaronson2005limitsefficientcomputationphysical,
      title={Limits on Efficient Computation in the Physical World}, 
      author={Scott Aaronson},
      year={2005},
      eprint={quant-ph/0412143},
      archivePrefix={arXiv},
      primaryClass={quant-ph},
      url={https://arxiv.org/abs/quant-ph/0412143}, 
}

@article{doi:10.1137/0206049,
author = {Gill, John},
title = {Computational Complexity of Probabilistic Turing Machines},
journal = {SIAM Journal on Computing},
volume = {6},
number = {4},
pages = {675-695},
year = {1977},
doi = {10.1137/0206049},
URL = { https://doi.org/10.1137/0206049},
eprint = {https://doi.org/10.1137/0206049}
}

@misc{dasilva2024demonstrationlogicalqubitsrepeated,
      title={Demonstration of logical qubits and repeated error correction with better-than-physical error rates}, 
      author={M. P. da Silva and C. Ryan-Anderson and J. M. Bello-Rivas and A. Chernoguzov and J. M. Dreiling and C. Foltz and F. Frachon and J. P. Gaebler and T. M. Gatterman and L. Grans-Samuelsson and D. Hayes and N. Hewitt and J. Johansen and D. Lucchetti and M. Mills and S. A. Moses and B. Neyenhuis and A. Paz and J. Pino and P. Siegfried and J. Strabley and A. Sundaram and D. Tom and S. J. Wernli and M. Zanner and R. P. Stutz and K. M. Svore},
      year={2024},
      eprint={2404.02280},
      archivePrefix={arXiv},
      primaryClass={quant-ph},
      url={https://arxiv.org/abs/2404.02280}, 
}

@article{PhysRevLett.128.110504,
  title = {Calibrated Decoders for Experimental Quantum Error Correction},
  author = {Chen, Edward H. and Yoder, Theodore J. and Kim, Youngseok and Sundaresan, Neereja and Srinivasan, Srikanth and Li, Muyuan and C\'orcoles, Antonio D. and Cross, Andrew W. and Takita, Maika},
  journal = {Phys. Rev. Lett.},
  volume = {128},
  issue = {11},
  pages = {110504},
  numpages = {7},
  year = {2022},
  month = {Mar},
  publisher = {American Physical Society},
  doi = {10.1103/PhysRevLett.128.110504},
  url = {https://link.aps.org/doi/10.1103/PhysRevLett.128.110504}
}

@article{Knill_2005, title={Quantum computing with realistically noisy devices}, volume={434}, DOI={10.1038/nature03350}, number={7029}, journal={Nature}, author={Knill, E.}, year={2005}, month={Mar}, pages={39–44}}
\end{document}